\documentclass[10pt,conference]{IEEEtran}
\usepackage{epsfig,setspace,amsmath,epsf,amssymb,bm,theorem,cite,graphicx, epstopdf,algorithm,algpseudocode,float,color,mathtools}
\usepackage[table,xcdraw]{xcolor}

\IEEEoverridecommandlockouts
\allowdisplaybreaks

\begin{document}

\title{Timely Multi-Goal Transmissions With an Intermittently Failing Sensor}

\author{Ismail Cosandal \qquad Sennur Ulukus\\
\normalsize Department of Electrical and Computer Engineering\\
\normalsize University of Maryland, College Park, MD 20742\\
\normalsize  \emph{ismailc@umd.edu} \qquad \emph{ulukus@umd.edu}}

\maketitle

\begin{abstract}
A sensor observes a random phenomenon and transmits updates about the observed phenomenon to a remote monitor. The sensor may experience intermittent failures in which case the monitor will not receive any updates until the sensor has recovered. The monitor wants to keep a timely view of the observed process, as well as to detect any sensor failures, using the timings of the updates. We analyze this system model from a \emph{goal-oriented} and \emph{semantic} communication point of view, where the communication has multiple goals and multiple meanings/semantics. For the first goal, the performance is quantified by the age of information of the observed process at the monitor. For the second goal, the performance is quantified by the probability of error of the monitor's estimation of the sensor's failure status. Each arriving update packet brings both an information update and an indication about the sensor's status. The monitor estimates the failure status of the sensor by using the timings of the received updates. This estimation is subject to error, since a long period without any update receptions may be due to a low update rate or a failure of the sensor. We examine the trade-off between these two goals. We show that the probability of error of estimating a sensor failure decreases with increased update rate, however, the age of information is minimized with an intermediate update rate (not too low or high).
\end{abstract}

\section{Introduction}
While initial studies on \emph{semantic} or \emph{goal-oriented} communications date back decades \cite{carnap1952}, it now presents itself as an idea whose time has finally come, thanks to envisioned applications in the next generation 6G wireless \cite{sagduyu2023multireceiver,sagduyu2023age,strinati2021sem11, yang2022sem1, qin2021sem12, lu2022semantics, gunduz2023}. Different from traditional communications, semantic communications prioritizes the semantic meaning, or the aim, of the message rather than successfully transmitting each and every bit of the message. Thus, the performance of communication is measured by how much the message has  served the goal/aim of the transmission \cite{lu2022semantics,getu-survey}. Bilingual evaluation understudy (BLEU) \cite{xie2023sem1250}, the age of information (AoI) \cite{uysal2022sem165}, and their variations \cite{maatouk2020} are the most frequently used performance metrics for this purpose. 

Further, the traditional approach assumes that each transmission conveys a single information and serves a single goal, and the system performance is measured using different metrics on this single goal, e.g., for the communication goal, typical metrics are probability of error, throughput, and delay. Instead, our approach in this paper is to serve multiple goals/purposes with a single transmission. This approach was used previously, e.g., in \cite{yang2018} where every transmission carries an update packet and packet timings carry an independent information, and in \cite{ozel2022} where every transmission carries an update packet, packet timings carry an independent information, and transmissions carry information about the battery state of the transmitter. 

\begin{figure}[t]
    \centering
    \includegraphics[]{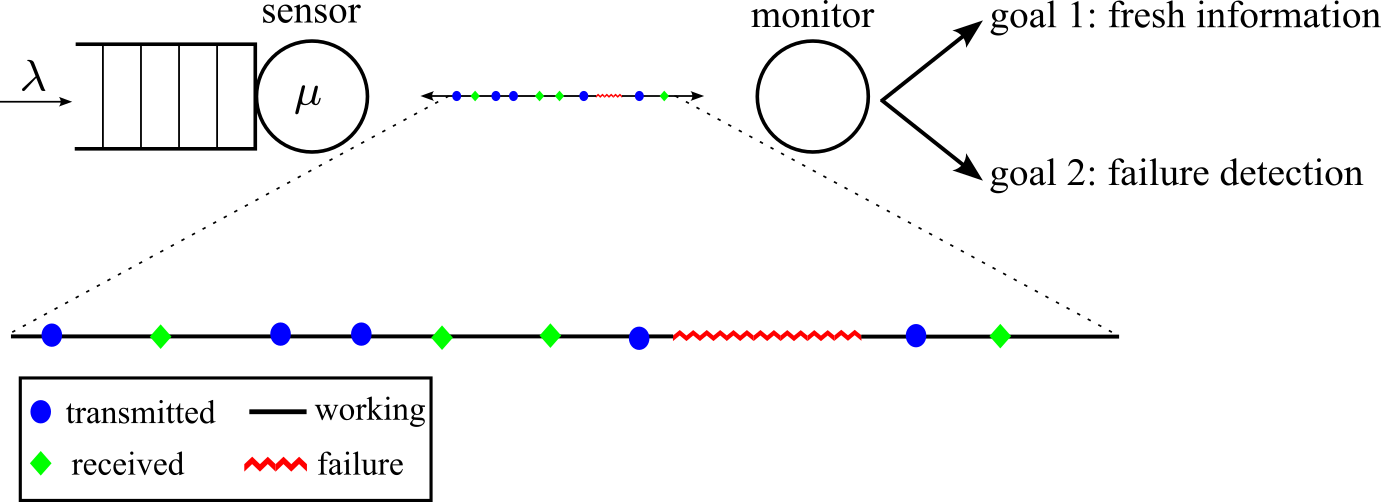} 
    \caption{The system model with an intermittently failing sensor and a monitor. When the sensor is operational, it generates updates with rate $\lambda$, sends them to the monitor through a queue with service rate $\mu$. Transmission stops during failures, and any packet queued up or being served is discarded. The monitor has two goals: obtaining fresh information and detecting any sensor failures from the incoming update packets and their timings.}
    \label{fig:sys}
    \vspace*{-0.4cm}
\end{figure}

In this paper, we consider a multi-goal system where a sensor observes a random phenomenon and sends update packets about it to a monitor; see Fig.~\ref{fig:sys}. The monitor has two goals: to keep a timely view of the phenomenon observed by the sensor and to keep an accurate view of whether the sensor is operational or experiencing a failure. Every incoming update packet serves both purposes in different ways: each packet carries update information about the observed phenomenon, and timings of the incoming packets carry information about the failure status of the sensor, e.g., if the receiver has not received an update for a long time, it may be an indication that the sensor has experienced a failure.

\begin{figure*}[t]
    \centering
    \includegraphics[width=1.9\columnwidth]{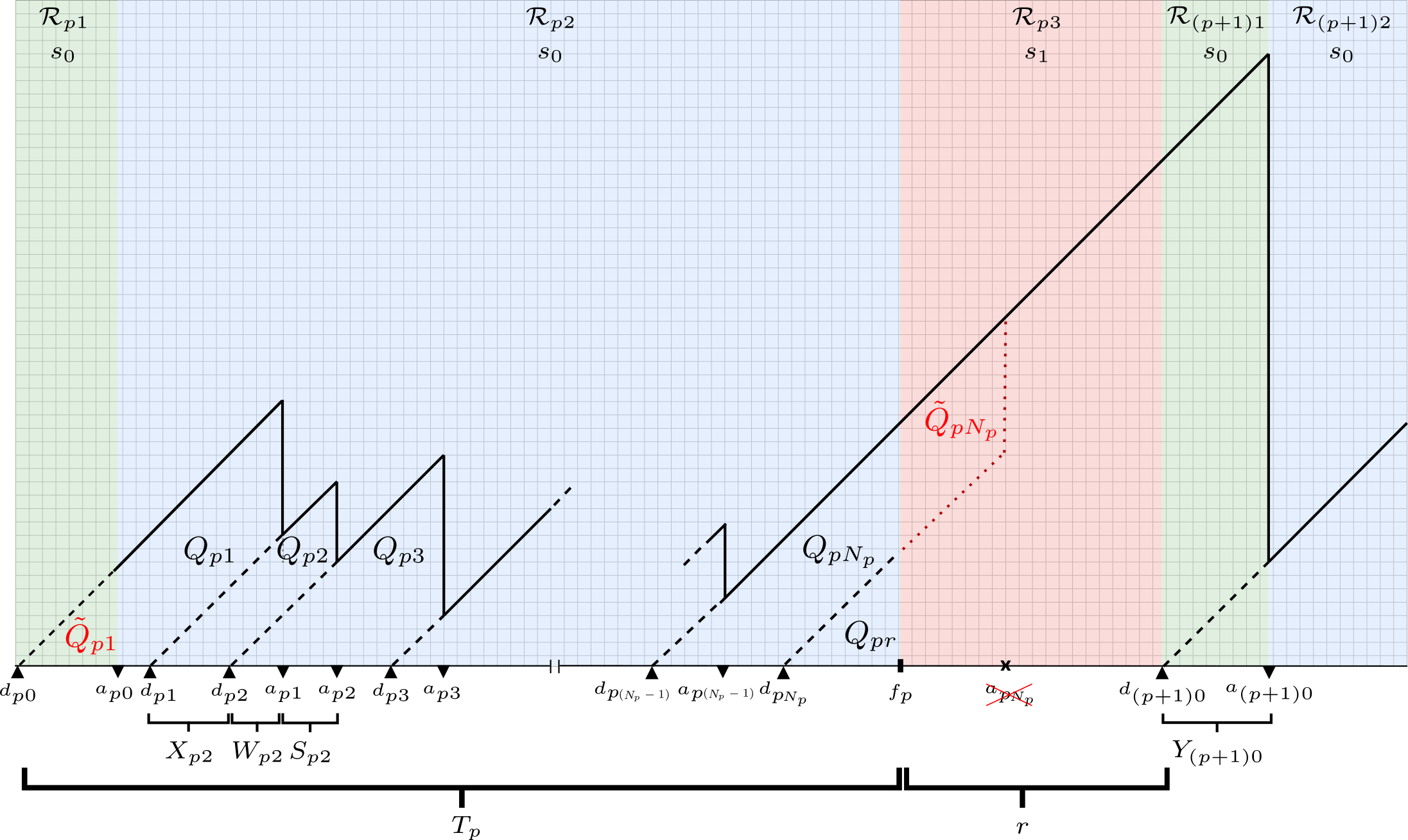} 
    \vspace*{-0.3cm}
    \caption{An example of age of information changes in the first period.}
    \label{fig:aoi}
    \vspace*{-0.3cm}
\end{figure*}

For the second goal, \emph{the failure status of the sensor}, we use the timings of the incoming packets at the receiver to estimate the status of the sensor using the maximum aposteriori probability (MAP) rule. We use the fact that if the sensor is sending updates with a rate $\lambda$, and if the service rate of the queue $\mu$ is larger than $\lambda$, then the packets should be exiting the queue at a rate $\lambda$. Thus, the hypothesis testing problem at the receiver is that packet timings are either exponential with rate $\lambda$ (sensor is working) or not. If the monitor estimates the sensor status as working while it is not, or vice versa, we have an error event. The error rate, which is the time averaged absolute difference between the estimated and actual sensor status is used to measure the performance of the estimate \cite{bastopcu2021, maatouk2020}.

For the first goal, the \emph{freshness of information}, we use the AoI as the metric \cite{yates-survey, sun-survey, kosta-survey} which is the time difference between the system time and the generation time of the last received data at the monitor. While AoI has been studied widely for different scenarios \cite{yates2012, yates2019, kam2018, chen2016}, ours is the first work that includes the failure and recovery process of the observing sensor in the AoI literature.

\section{System Model}
We consider a sensor node which observes a random phenomenon, generates status updates, and sends them to a monitor by using a first-come first-served (FCFS) queue until a failure happens. When a sensor failure happens, the sensor stops generating and sending updates, and remains idle until recovery. After the recovery, the sensor starts to generate updates as usual. We denote the normal operational status by $s_0$ and the failure status by $s_1$ during which recovery happens.

As illustrated with an example in Fig.~\ref{fig:aoi}, the $p$th period starts with the first update generation after the $(p-1)$th failure. In the $p$th period, the departure (from the transmitter) time of the $k$th update is denoted by $d_{pk}$ and the arrival time of the $k$th update packet to the monitor is denoted by $a_{pk}$. At time $f_p$ a failure happens and the sensor enters the state $s_1$, thus, the last update generated before the failure is never received. The recovery process ends when the sensor generates a new update, and the next period starts at that time. 

We consider an $M/M/1$ queueing system meaning that the inter-generation time between two updates in period $p$ is $X_{pk}=d_{pk}-d_{p(k-1)}$ and the service time of the corresponding update $S_{pk}$ are independent and identically distributed (iid) exponential random variables with $\mathbb{E}[X] = \frac{1}{\lambda}$ and $\mathbb{E}[S]=\frac{1}{\mu}$, respectively. Thus, the service utilization ratio is $\rho=\frac{\lambda}{\mu}$. Service duration of the $k$th packet is $Y_{pk}=a_{pk}-d_{pk}$, which is the sum of the waiting time $W_{pk}$ and the service time $S_{pk}$. Finally, $T_{p}$ is the duration before a failure happens in the $p$th period, and $r$ is the recovery time which is a constant. 

The monitor has two goals: to receive the freshest information and to detect sensor failures. For the first goal, AoI is the performance metric we use, 
\begin{align}
    \Delta(t)=t-\max\{d_{pk} \mid t\geq a_{pk} \}.
\end{align} 
For the second goal, the probability of error is the performance metric we use, which can also be expressed as the expected absolute difference between the estimated status $\hat{s}(t)$ and the actual status $s(t)$,
\begin{align}
 E=\mathbb{E}[|\hat{s}(t)-s(t)|]. 
\end{align} 

In a full period, we have three regions, see Fig.~\ref{fig:aoi}:  
\begin{itemize}
\item $\mathcal{R}_{p1}$: This region starts with the generation of an update at $d_{p0}$, and lasts for the first service time in $a_{p0}$. Thus, its duration is $Y_{p0}$. Even though no update is received during this region, the failure status is $s_0$.

\item $\mathcal{R}_{p2}$: This region starts after the completion of the service of the packet generated after the recovery at $a_{p0}$, and lasts until a failure happens again. Thus, its duration is $f_{p}-a_{p0}$, which is equal to $T_p-Y_{p0}$. In this region, the failure status is $s_0$. We denote the number of updates generated in this region as $N_{p}=\max \{ k \ | \ d_{pk} \leq f_{p} \}$.

\item $\mathcal{R}_{p3}$: This is the region that starts with a failure at $f_{p}$, and lasts until the recovery at $d_{(p+1)0}$. The duration of this region is deterministic, and is equal to $r$. In this region, the failure status is $s_1$, and AoI increases monotonically.
\end{itemize}

\section{Analysis}
In this section, we derive analytical expressions for both metrics for an intermittently failing sensor. During these calculations, we use the following three assumptions:
\begin{itemize}
    \item The occurrence of a failure event for an infinitesimal time duration is a constant,
    \begin{align}
        \lim_{u \to 0} \frac{\mathbb{P}(t<T<t+u \mid t>T)}{u}=\nu.
    \end{align} 
    Thus, the duration before failures $T_{1},T_{2},\dots,T_{p}$ are iid exponential random variables with $\mathbb{E}[T]=\frac{1}{\nu}$.
    \item The recovery process starts right after the failure, independent of the detection of the failure by the sensor, and it lasts $r$ seconds, where $r$ is deterministic.
    \item There is at least one successful service after the recovery, in other words, $T_{p} > Y_{p0}$. This assumption excludes consecutive failures and guarantees that each period includes three regions (see green, blue, pink regions in Fig.~\ref{fig:aoi}).
\end{itemize}

In the following, we generally omit the subscript for the period, i.e., $p$ in the subscripts, while calculating the average values over a single period. However, throughout the analysis, we calculate the time average of metrics over multiple periods.

\subsection{Sensor Failure Status Detection and the Error Rate}
First, we derive the optimum detection method for the failure status of the sensor using the incoming packet timings. Then, we evaluate the performance of the proposed detection method via probability of error of detection. During the calculations, we utilize Burke's theorem \cite{burke1956, cassandras2008book}, which states that for an $M/M/1$ queue, the departures form a Poisson process with the same parameter as the arrivals ($\lambda$ in this case), under the stability condition for the queue, i.e., $\lambda<\mu$, equivalently, $\rho<1$. That is, so long as the sensor is taking update samples with exponential inter-sampling rate $\lambda$, and these updates are transmitted via a queue of exponential service rate $\mu$ where $\lambda <\mu$, then the update packets arrive at the monitor as a Poisson process with an exponential inter-arrival rate $\lambda$.

We denote the $k$th interval of received updates at the monitor as $I_k=a_{k+1}-a_k$, which has exponential distribution with $\mathbb{E}[I]=\frac{1}{\lambda}$. In addition, we denote the time elapsed since the last received update with $z(t)$, where
\begin{align}
z(t) =t-\max\{a_k \mid a_k\leq t\}.
\end{align}
Here, we exclude the region $\mathcal{R}_1$, and derive the optimum detection rule in regions $\mathcal{R}_2$ and $\mathcal{R}_3$ based on observing $z(t)$. 

In region $\mathcal{R}_2$, $z(t)$ linearly increases until a new update comes or the region ends with a sensor failure. In order for $z(t)$ to reach a value $\zeta$, neither of the two aforementioned events should occur for $\zeta$. We calculate this probability as,
\begin{align}
   \mathbb{P}(z(t)>\zeta \mid \mathcal{R}_2)=&1-F_Z(\zeta \mid \mathcal{R}_2)\nonumber\\
   =&(1-F_T(\zeta))(1-F_I(\zeta))\nonumber \\ 
   =&e^{-(\lambda+\nu)\zeta},
\end{align}
where $F_V(v)$ is the cumulative distribution function (cdf) of the random variable $V$. Then, pdf of $z(t)$ is
\begin{align}
    f_Z(z(t) \mid \mathcal{R}_2)=(\lambda+\nu) e^{-(\lambda+\nu)z(t)}.
\end{align}

In region $\mathcal{R}_3$, the value of $z(t)$ starts from $z(f)$ and linearly increases until $z(f)+r$. Thus, inside of the time interval of $[f,f+r]$, $z(t)$ can take any value in $[z(f),z(f)+r]$ with equal likelihood. Because of this linear relation, the conditional random variable $z(t) | \mathcal{R}_3$ can be expressed as a sum of $z(f)$ and $u$ as,
\begin{align}
    z(t) \mid \mathcal{R}_3 = z(f) + u,
\end{align}
where $z(f)$ has the same distribution as $z(t) | \mathcal{R}_2$, and $u$ is a uniform random variable in $[0, r]$. As the pdf of sum of two independent random variables is the convolution their pdfs, we obtain the pdf of $z(t)$ in region $\mathcal{R}_3$ as
\begin{align}
\!\!\!\!f_Z(z(t)\mid \mathcal{R}_3)&= (\lambda+\nu)e^{-(\lambda+\nu)z(t)}\int_0^{\min(r,z(t))}\frac{e^{(\lambda+\nu)u}}{r}du \nonumber \\
&=\begin{cases} \frac{1-e^{-(\lambda+\nu)z(t)}}{r}, & 0\leq z(t) < r \\ \frac{e^{-(\lambda+\nu)z(t)}(e^{(\lambda+\nu) r}-1)}{r}, & z(t)\geq r \end{cases} \!\!
\end{align}
By using these conditional pdfs and the ratio of apriori state probabilities $\frac{\mathbb{P}(s_1)}{\mathbb{P}(s_0)}={r\nu}$, we drive the decision rule as follows. 

First, when $0\leq z(t)< r$, we start with
\begin{align}
 f_Z(z(t)|\mathcal{R}_2)\mathbb{P}(s_0) \mathop{\gtreqless}_{s_1}^{s_0} f_Z(z(t)|\mathcal{R}_3)\mathbb{P}(s_1) 
\end{align}
which is equivalent to
\begin{align}
(\lambda+\nu) e^{-(\lambda+\nu)z(t)} \mathop{\gtreqless}_{s_1}^{s_0} \frac{1-e^{-(\lambda+\nu) z(t)}}{r} {r\nu} 
\end{align}
which gives
\begin{align}
\frac{\log\left(\frac{\lambda}{\nu}+2\right)}{\lambda+\nu}\mathop{\gtreqless}_{s_1}^{s_0} z(t). \label{eq:tau1}   
\end{align}
Let us define the threshold $\tau=\frac{\log\left(\frac{\lambda}{\nu}+2\right)}{\lambda+\nu}$. The decision rule in (\ref{eq:tau1}) indicates that, if the monitor has not received an update packet within the last $\tau$ seconds since the last update packet, it estimates the status of the sensor as it is in failure. Whereas if the monitor receives an update within $\tau$ seconds since the last update, it estimates the status of the sensor as it is operational. 

Next, when $z(t)>r$, we start with
\begin{align}
f_Z(z(t)\mid \mathcal{R}_2) \mathbb{P}(s_0) \mathop{\gtreqless}_{s_1}^{s_0} f_Z(z(t)\mid \mathcal{R}_3) \mathbb{P}(s_1) 
\end{align}
which is equivalent to
\begin{align}
(\lambda+\nu) e^{-(\lambda+\nu)z(t)} \mathop{\gtreqless}_{s_1}^{s_0} \frac{e^{-(\lambda+\nu) z(t)}(e^{(\lambda+\nu) r}-1)}{r} r\nu
\end{align}
which gives
\begin{align}
\frac{\log\left(\frac{\lambda}{\nu}+2\right)}{\lambda+\nu}\mathop{\gtreqless}_{s_1}^{s_0} r.
\label{eq:tau2}
\end{align}
Note that we obtained the same threshold $\tau$ on the left hand side of the inequality in (\ref{eq:tau2}), and also that this decision rule does not depend on the value of $z(t)$. 

Decision rules in (\ref{eq:tau1}) and (\ref{eq:tau2}) when put together say the following: If the system parameters $\lambda$, $\nu$ and $r$ are such that $\tau>r$, then the monitor should always estimate the sensor status as operational. This happens, for example, when $r$ is small (sensor recovers quickly) and $\nu$ is small (failures happen rarely). This constitutes a degenerate case. In practically relevant cases, the threshold $\tau$ is smaller than $r$ to yield a non-degenerate decision rule for the monitor. Then, by combining \eqref{eq:tau1} and \eqref{eq:tau2}, we obtain the decision rule as
\begin{align}
    \hat{s}(t)=\begin{cases}
    s_0, & \text{if} \ \tau \geq r \ \text{irrespective of} \ z(t),\\
    s_0, & \text{if} \ z(t) \leq \tau \ \text{and} \ \tau < r, \\
    s_1, & \text{if} \ z(t) > \tau \ \text{and} \ \tau < r.
    \end{cases} \label{eq:tau}
\end{align}

Finally, we calculate the \emph{probability of error} for the monitor's estimate of the sensor's status when $\tau<r$, as
\begin{align}    
\!E=&\mathbb{E}\left[\left|\hat{s}(t)-s(t)\right|\right] \nonumber \\
=&\mathbb{E}\left[\mathbf{1}_{\{z(t)\geq\tau\}} \mid s_0\right]\mathbb{P}(s_0)+\mathbb{E}\left[\mathbf{1}_{\{z(t)<\tau\}}\mid s_1\right]\mathbb{P}(s_1) \nonumber  \\
=& \frac{1}{1+r\nu} \int_\tau^\infty (\lambda+\nu)e^{-(\lambda+\nu)z}dz \nonumber \\
 & +\frac{r \nu}{1+r\nu}\int_0^\tau \frac{1-e^{-(\lambda+\nu)z}}{r}dz \nonumber  \\
=&\frac{1}{1+r\nu} \frac{\nu}{\lambda+2\nu}+\frac{\nu}{1+r\nu}\frac{\log\left(\frac{\lambda}{\nu}+2\right)+\frac{\nu}{\lambda+2\nu}-1}{\lambda+\nu}. \label{eq:error-rate}
\end{align}

This result in (\ref{eq:error-rate}) shows that both false positive rate and false negative rate decrease with increased $\lambda$ and $\nu$. In other words, receiving frequent updates or experiencing rare failures decreases the error rate. In addition, even though we observe from (\ref{eq:tau1}), (\ref{eq:tau2}) and (\ref{eq:tau}) that the recovery duration $r$ has no effect on the threshold $\tau$ or the decision rule, it decreases the error rate in (\ref{eq:error-rate}) by increasing the total duration of a period.

\subsection{Age of Information Calculation}

One of our assumptions is that each period includes three regions. Thus, to calculate the average age, we analyze the average AoI in each region.

\paragraph{Average AoI in Region  $\mathcal{R}_2$} This region is similar to FCFS model in \cite{yates2012} except that it has finite duration. Using the definition of time-average, we have the average AoI as 
\begin{align} 
\Delta_{\mathcal{R}_2}= \frac{1}{T-Y_0} \int_{a_0}^{f} \Delta(t)dt, 
\end{align} 
which, geometrically interpreted from Fig.~\ref{fig:aoi}, is given as
\begin{align} 
\Delta_{\mathcal{R}_2}= \frac{Q_r-\Tilde{Q}_{1}-\Tilde{Q}_{N}+\sum_{n=1}^{N}Q_n}{T-Y_0}, \label{eq1} 
\end{align} 
where $Q_n$ is the area of the trapezoid between $d_{n-1}$ and $d_{n}$, 
\begin{align} 
Q_n=\frac{X_n^{2}}{2}+X_nY_n. \label{eq2} 
\end{align} 
Furthermore, as shown in Fig.~\ref{fig:aoi}, areas exceeding from $\mathcal{R}_2$ are denoted by $\Tilde{Q}_1$ and $\Tilde{Q}_N$, and there is a residual area $Q_r$ between regions $\mathcal{R}_2$ and  $\mathcal{R}_3$. Inserting \eqref{eq1} into \eqref{eq2}, we obtain the average AoI in region $\mathcal{R}_2$ as
\begin{align}
\!\!\Delta_{\mathcal{R}_2}=\frac{Q_r\!-\!\Tilde{Q}_{1}\!-\!\Tilde{Q}_{N}}{T\!-\!Y_0} \!+\! \frac{N}{T\!-\!Y_0}\frac{1}{N} \sum_{n=1}^{N}\left( X_nY_n\!+\!\frac{X_n^{2}}{2}\right).  \!  \label{eq3}  
\end{align}

Note that \eqref{eq3} will be equivalent to \cite[Eqn.~(5)]{yates2012} as $N \to \infty$. However, because the time interval $(T-Y_0)$ is finite, we cannot use the further steps in \cite{yates2012}. Specifically, we cannot say that the first term in \eqref{eq3} goes to zero, and also, it is not guaranteed that the system reaches a steady state. Recall that the service duration is the sum of the service duration $Y_n$ and the waiting time in the queue $W_n$. Additionally, the waiting time in the queue is $W_n=\max\{0,Y_{n-1}-X_n\}$. Thus, $Y_n$ depends on the previous arrival and service times. Note that while each waiting time has different statistical characteristics, when the system reaches its steady state, the distribution of $W$ approaches the exponential distribution with the expected value  $\frac{\rho}{\mu(1-\rho)}$ \cite{yates2012}. However, in our system model, it is not guaranteed that the system will reach a steady state, because a failure may happen before the system reaches its steady state. 

To obtain an analytical expression, we note that as $\mathbb{E}[T]$ increases (i.e., as the expected time in the non-failure state increases, equivalently as the probability of failure decreases), $N$ increases making the aforementioned assumptions valid. Then, we have the following approximation
\begin{align}  
\mathbb{E} [ \Delta_{\mathcal{R}_2} ]  = \frac{1}{\mu} \left( 1 + \frac{1}{\rho}+\frac{\rho^2}{1-\rho} \right)=\Delta_{M/M/1}.  
\end{align} 
 
\paragraph{Average AoI in Region $\mathcal{R}_3$} As seen from the Fig.~\ref{fig:aoi}, the instantaneous age starts with the value $\Delta(f)$ and increases to $\Delta (f)+r$ in the duration of $r$, which gives average AoI as
\begin{align} 
\Delta_{\mathcal{R}_2}= \Delta (f) +\frac{r}{2}. 
\end{align} 
Furthermore, because the time before failure $T$ is independent of $X_n$ and $Y_n$, we have $\mathbb{E}[\Delta (f)]=\mathbb{E}[\Delta_{\mathcal{R}_2}]$. Then, we have
\begin{align} 
\mathbb{E}[\Delta_{\mathcal{R}_3}] = \Delta_{M/M/1} + \frac{r}{2}.
\end{align} 

\paragraph{Average AoI in Region $\mathcal{R}_1$} Similar to $\mathcal{R}_3$, the average age in this region can also be expressed as 
\begin{align} 
\Delta_{\mathcal{R}_1}= \Delta (d_0) +\frac{Y_0}{2}, 
\end{align} 
with the equality $\mathbb{E}[\Delta (d_0)]= \mathbb{E}[ \Delta_{\mathcal{R}_2} ]+r$. Then, we have
\begin{align} 
\mathbb{E}[ \Delta_{\mathcal{R}_1} ] =& \mathbb{E} [ \Delta_{\mathcal{R}_2} ] + r + \frac{\mathbb{E}[Y]}{2} = \Delta_{M/M/1} + r + \frac{1}{2\mu}.
\end{align} 

\paragraph{Overall Average AoI} Now, we calculate the time average of AoI by taking an average over $P$ periods. Because of our assumption $T_p\geq Y_{p0}$, each period consists of the above three regions, and we express the time average over $P$ periods when $P$ goes to the infinity as
\begin{align} 
\mathbb{E}[\Delta]=& \lim_{P \to \infty}\frac{ \sum_{p=1}^{P} \left( \Delta_{\mathcal{R}_{p1}} \ell_{p1} + \Delta_{\mathcal{R}_{p2}} \ell_{p2}+ \Delta_{\mathcal{R}_{p3}} \ell_{p3} \right) }{\sum_{p=1}^{P} (T_p+r) }\nonumber \\
=& \lim_{P \to \infty}\frac{ \frac{1}{P}\sum_{p=1}^{P} \left( \Delta_{\mathcal{R}_{p1}} \ell_{p1} +\ \Delta_{\mathcal{R}_{p2}} \ell_{p2}+ \Delta_{\mathcal{R}_{p3}} \ell_{p3} \right) }{\frac{1}{P}\sum_{p=1}^{P} (T_p+r) }\nonumber \\
=& \frac{\mathbb{E}\left[   \Delta_{\mathcal{R}_{1}} \ell_{1} + \Delta_{\mathcal{R}_{2}} \ell_{2}+ \Delta_{\mathcal{R}_{3}} \ell_{3} \right]}{\mathbb{E}\left[T+r\right] }, \label{eq:final-age}
\end{align} 
where $\ell_{p1}=Y_{p0}$, $\ell_{p2}=T_p-Y_{p0}$ and $\ell_{p3}=r$ are the durations of the three regions. By using our previous calculations and the fact that $T$ and $Y$ are independent random variables, we evaluate (\ref{eq:final-age}) as
\begin{align} 
\!\!\!\mathbb{E}[\Delta]=&\frac{ \mathbb{E} \left[ \Delta_{M/M/1} \left( T+r \right) +\frac{r^2}{2} +Yr+\frac{Y^2}{2}\right] }{\mathbb{E}[T+r]}\nonumber\\
=&\Delta_{M/M/1}+\frac{ \mathbb{E}\left[  \frac{r^2}{2}+Yr +\frac{Y^2}{2}\right]  }{{\mathbb{E}[T+r]}} \nonumber  \\
=&\frac{1}{\mu} \left( 1 + \frac{1}{\rho}+\frac{\rho^2}{1-\rho} \right)\!+\!\left(\frac{r^2}{2}+\frac{r}{\mu}+\frac{1}{\mu^2} \right)\frac{\nu}{1+r\nu}.\!\!
\label{eq:Eaoi}
\end{align}

Note that the only first term in \eqref{eq:Eaoi} depends on the service utilization rate $\rho$, thus the same $\rho$ in \cite{yates2012} minimizes it. The remaining term shows that the average age increases with the recovery duration $r$ and the failure frequency $\nu$.

\begin{figure}[t]
    \centering
    \includegraphics[width=\columnwidth]{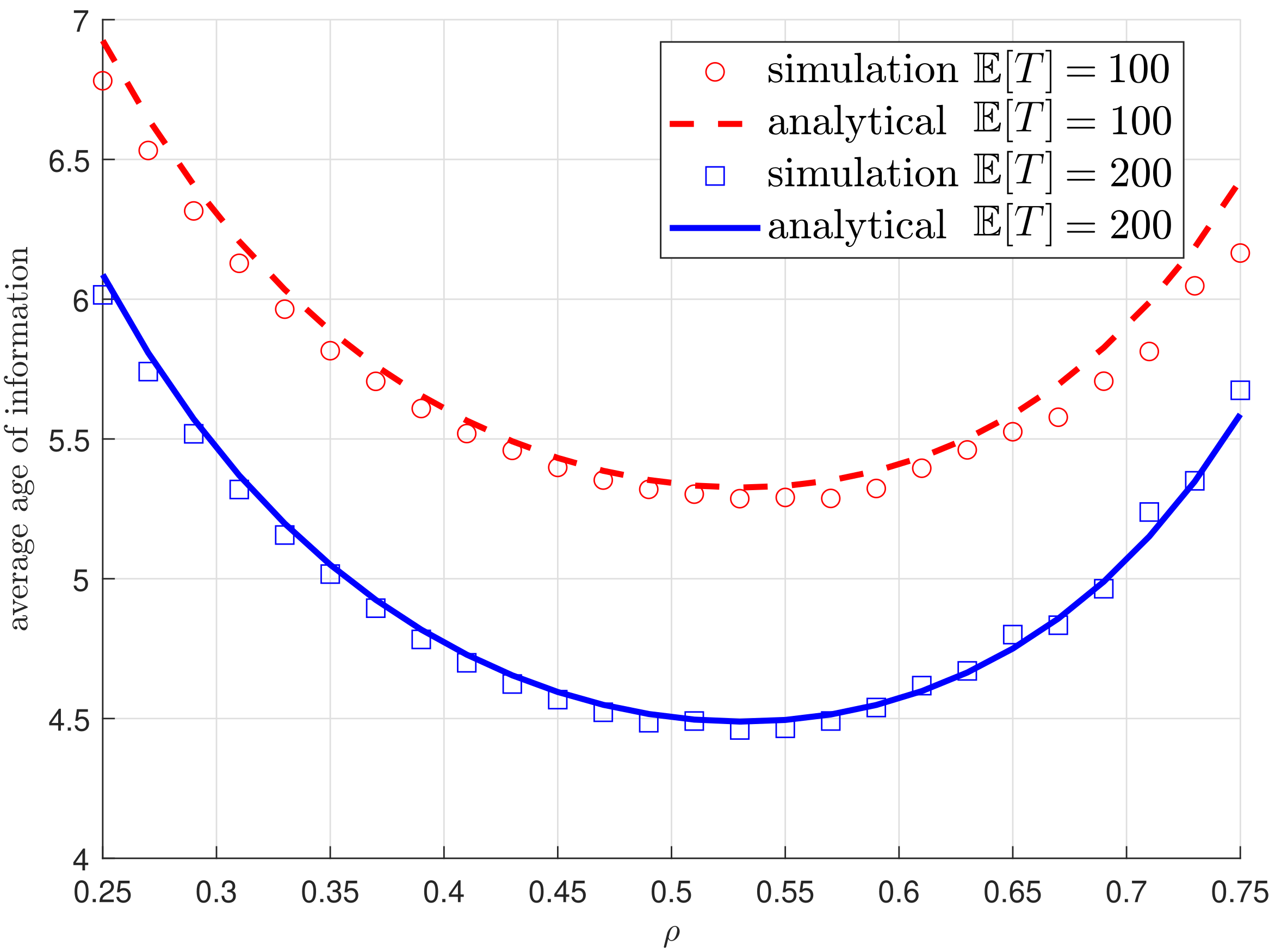}
    \caption{Analytical and simulation results for the AoI for different service utilization ratios $\rho$.}
    \label{fig:age}
    \vspace*{-0.4cm}
\end{figure}

\begin{figure}[t]
    \centering
    \includegraphics[width=\columnwidth]{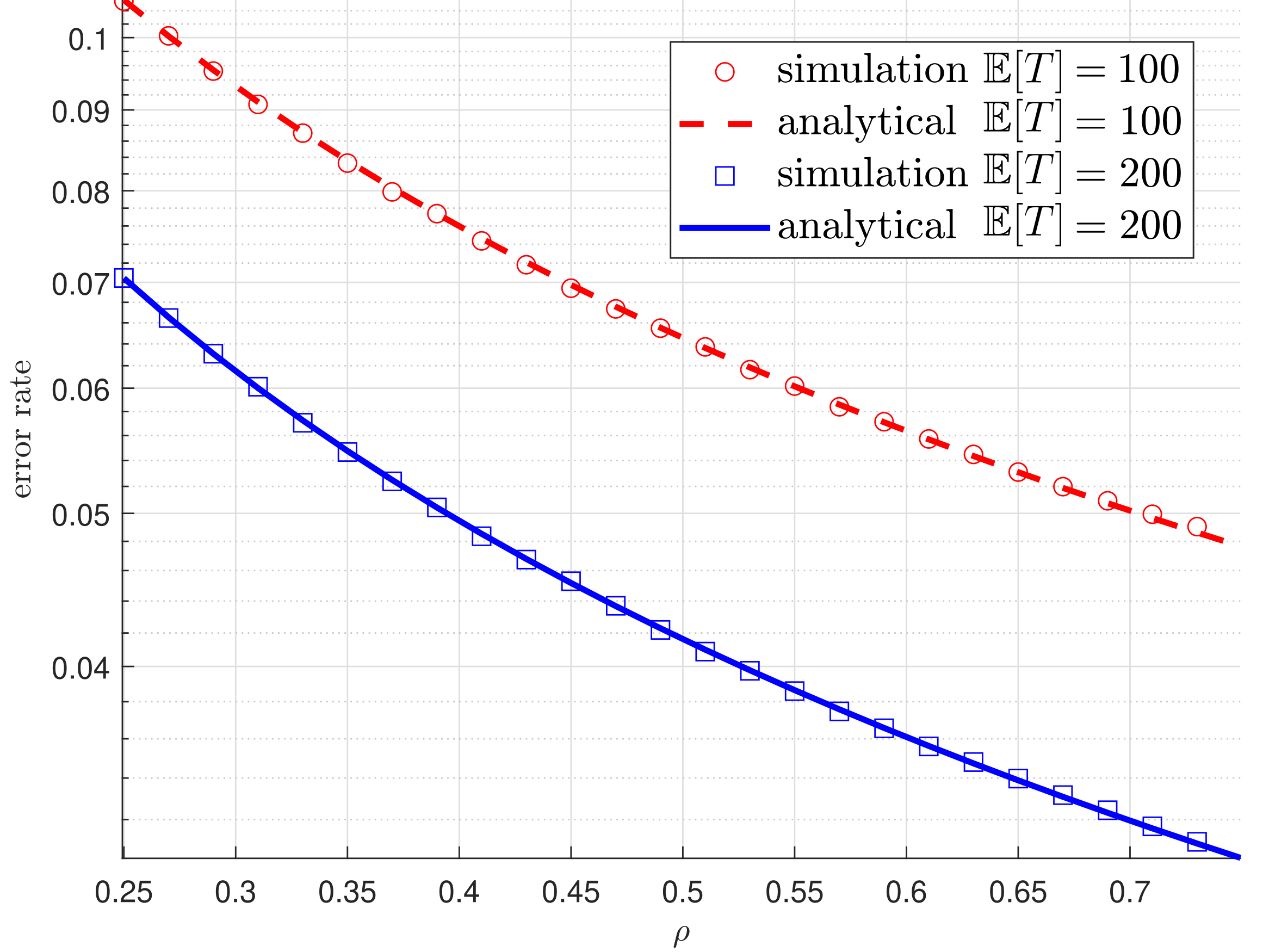} 
    \caption{Analytical and simulation results for the error rate for different service utilization ratios $\rho$.}
    \label{fig:er}
\end{figure}

\begin{figure}[t]
    \centering
    \includegraphics[width=\columnwidth]{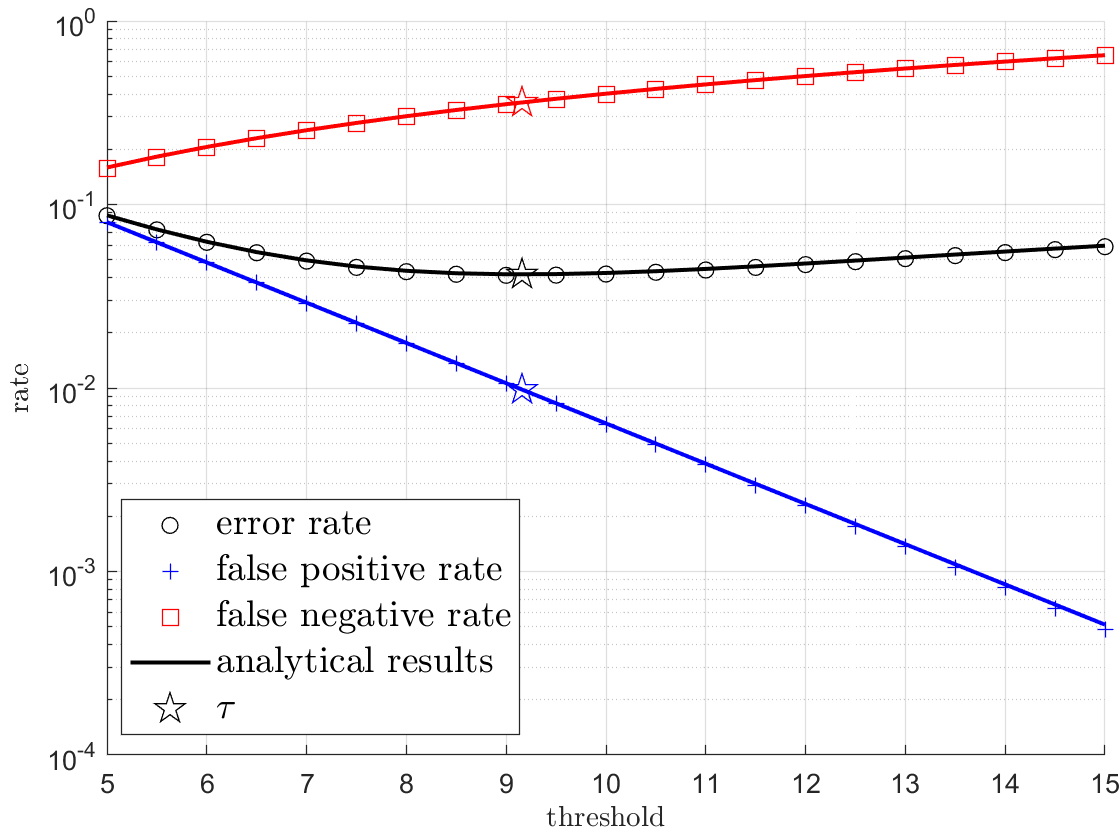}
    \caption{The error rate in the estimation of the sensor failure status when different threshold values are used. Here, $\tau$ shows the threshold value in \eqref{eq:tau1}, \eqref{eq:tau2} and used in \eqref{eq:tau}.}
    \label{fig:ext}
    \vspace*{-0.4cm}
\end{figure}

\section{Numerical Results}

In this section, we simulate the studied system model, and compare simulation results with the analytical results obtained in this paper. In these simulations, the averages of performance metrics are obtained over $P=10^5$ periods. Unless otherwise mentioned explicitly, the simulation parameters are selected as $\mathbb{E}[T]=\frac{1}{\nu}=200$, $\mu=1$, $\lambda=\frac{1}{2}$ (thus, $\rho=\frac{1}{2}$), $r=20$, and the state estimation is done by the decision rule in \eqref{eq:tau}. 

In Fig.~\ref{fig:age}, analytical and simulation results are shown for different $\mathbb{E}[T]$ and $\rho$ values. We observe a general agreement between the analytical and simulation results, and the agreement gets better for the larger $\mathbb{E}[T]$ value. The reason for this is that our calculations are based on the assumption that the system reaches a steady state in $\mathcal{R}_2$, and as the number of received updates increases before a failure, the system gets closer to the steady state. Further, in low $\mathbb{E}[T]$, the probability that the system enters $\mathcal{R}_3$ without entering $\mathcal{R}_2$ increases.

Fig.~\ref{fig:er} shows the analytical and simulation results for the error rate. We observe a close agreement between the analytical and simulation results. Fig.~\ref{fig:ext} shows the error rate when different threshold values are used for the estimation of the sensor failure status. For the parameters of this simulation setting, the threshold found in (\ref{eq:tau1}), (\ref{eq:tau2}) and used in (\ref{eq:tau}) is $\tau=\frac{\log\left(\frac{\lambda}{\nu}+2\right)}{\lambda+\nu}=9.16$, which is shown with a star in Fig.~\ref{fig:ext}, and it minimizes the error rate.

\begin{figure}[t]
    \centering
    \includegraphics[width=\columnwidth]{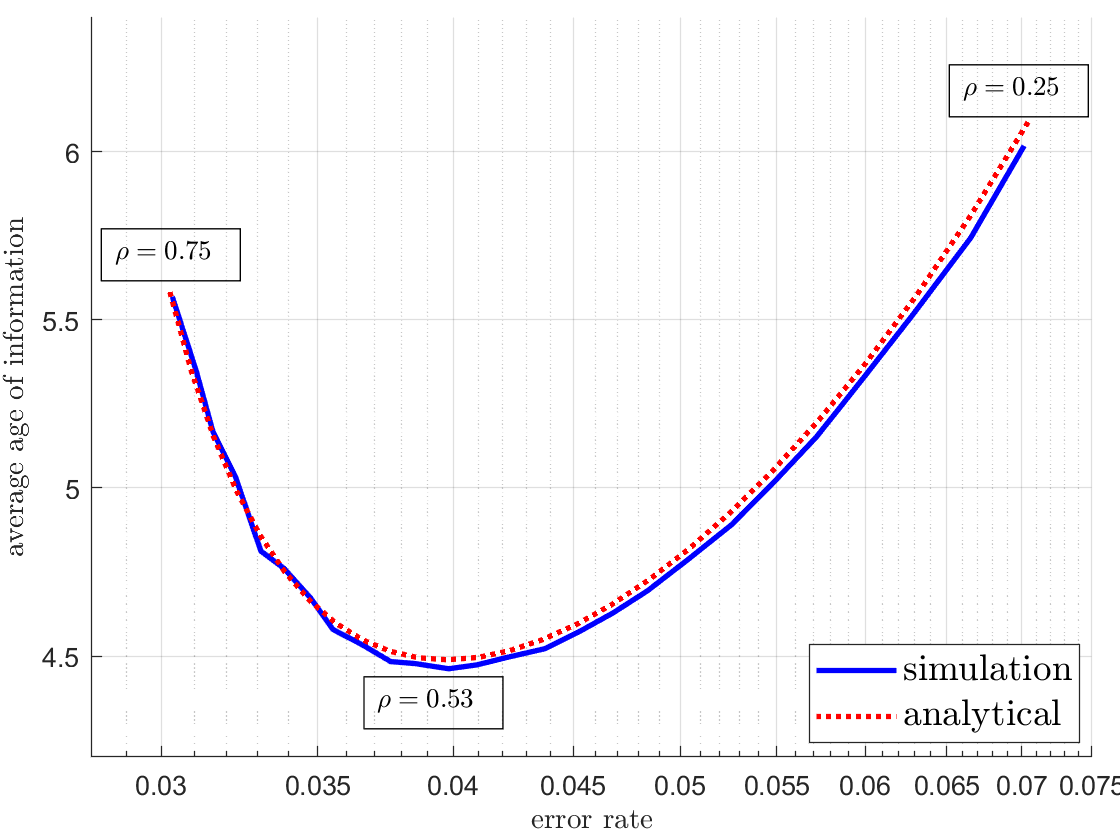}
    \caption{The trade-off between AoI for goal~1, and error rate for goal~2 for different $\rho$ values when $\mathbb{E}[T]=200$ and $r=20$.}
    \label{fig:comp}
    \vspace*{-0.2cm}
\end{figure}

Fig.~\ref{fig:comp} shows the two metrics used for the two goals considered in this paper: error rate of estimating the sensor failure status and the AoI of the random phenomenon observed by the sensor at the monitor side. We plot these two metrics by spanning $\rho$ through all possible values. First, we observe that the analytical and simulation results are in close agreement. Second, we observe that while the same $\rho$ value that minimizes the AoI of the generic $M/M/1$ model in \cite{yates2012} minimizes the AoI here where we have failures and recoveries of the sensor, the error rate of the estimation of the sensor failure status decreases with $\rho$. Therefore, if there is a constraint on the error rate, the optimal $\rho$ may be different than $0.53$. Similarly, if there is an AoI constraint, $\rho$ should be chosen close to $0.53$, even if that makes the error rate higher.

\section{Conclusion and Discussion}
We analyzed a system model consisting of a sensor that intermittently experiences failures and recoveries while sending time-sensitive updates to a monitor. In this system model, each transmitted update serves two goals: AoI of the updates at the monitor and the error rate of the estimation of the failure status of the sensor. For the error rate, we first derived the optimal decision rule for the estimation of the failure status of the sensor and derived a closed-form expression for its performance. For the AoI, we derived an approximate analytical expression, which becomes more accurate as the number of arrived updates in each period increases. We observed that while the AoI is minimized when the service utilization ratio is close to $0.53$ as in the classical $M/M/1$ system, the error rate monotonically decreases with the service utilization ratio.

Several extensions of this work could be studied: First, one can consider the case where the recovery time is not a deterministic constant $r$ as assumed in this work but is a random variable. Second, one can consider the case where the sensor does not recover on its own (after a deterministic or a random time), but a repair mechanism should be dispatched once the failure status of the sensor is detected. In this case, there will be more interesting trade-offs between the AoI and the error rate, as undetected failures will increase the age. This suggests that increasing false positives may decrease the AoI even though it increases the error rate. On the other hand, if there is a cost for dispatching a repair team, then false positives will increase the system cost while decreasing the AoI. 

\bibliographystyle{unsrt}
\bibliography{bibl}

\begin{thebibliography}{10}

\bibitem{carnap1952}
R.~Carnap and Y.~Bar-Hillel.
\newblock An outline of a theory of semantic information.
\newblock {\em Research Laboratory of Electronics, Massachusetts Institute of
  Technology}, (247), October 1952.

\bibitem{sagduyu2023multireceiver}
Y.~E. Sagduyu, T.~Erpek, A.~Yener, and S.~Ulukus.
\newblock Multi-receiver task-oriented communications via multi-task deep
  learning.
\newblock Available online at arXiv:2308.06884.

\bibitem{sagduyu2023age}
Y.~E. Sagduyu, S.~Ulukus, and A.~Yener.
\newblock Age of information in deep learning-driven task-oriented
  communications.
\newblock In {\em IEEE Infocom}, June 2023.

\bibitem{strinati2021sem11}
E.~C. Strinati and S.~Barbarossa.
\newblock 6{G} networks: Beyond {S}hannon towards semantic and goal-oriented
  communications.
\newblock {\em Computer Networks}, 190:107930, May 2021.

\bibitem{yang2022sem1}
W.~Yang, H.~Du, Z.~Q. Liew, W.~Y.~B. Lim, Z.~Xiong, D.~Niyato, X.~Chi, X.~Shen,
  and C.~Miao.
\newblock Semantic communications for future internet: Fundamentals,
  applications, and challenges.
\newblock {\em IEEE Communications Surveys \& Tutorials}, 25(1), November 2022.

\bibitem{qin2021sem12}
Z.~Qin, X.~Tao, J.~Lu, W.~Tong, and G.~Y. Li.
\newblock Semantic communications: Principles and challenges.
\newblock Available online at arXiv:2201.01389.

\bibitem{lu2022semantics}
Z.~Lu, R.~Li, K.~Lu, X.~Chen, E.~Hossain, Z.~Zhao, and H.~Zhang.
\newblock Semantics-empowered communication: A tutorial-cum-survey.
\newblock Available online at arxiv:2212.08487.

\bibitem{gunduz2023}
D.~Gunduz, Z.~Qin, I.~E. Aguerri, H.~S. Dhillon, Z.~Yang, A.~Yener, K.~K. Wong,
  and C.~B. Chae.
\newblock Beyond transmitting bits: Context, semantics, and task-oriented
  communications.
\newblock {\em IEEE Journal on Selected Areas in Communications}, 41(1):5--41,
  January 2023.

\bibitem{getu-survey}
T.~M. Getu, G.~Kaddoum, and M.~Bennis.
\newblock Making sense of meaning: A survey on metrics for semantic and
  goal-oriented communication.
\newblock {\em IEEE Access}, 11:45456--45492, April 2023.

\bibitem{xie2023sem1250}
H.~Xie, Z.~Qin, G.~Y. Li, and B.~H. Juang.
\newblock Deep learning enabled semantic communication systems.
\newblock {\em IEEE Transactions on Signal Processing}, 41(1):170--185, January
  2023.

\bibitem{uysal2022sem165}
E.~Uysal, O.~Kaya, A.~Ephremides, J.~Gross, M.~Codreanu, P.~Popovski,
  M.~Assaad, G.~Liva, A.~Munari, B.~Soret, and K.~H. Johansson.
\newblock Semantic communications in networked systems: A data significance
  perspective.
\newblock {\em IEEE Network}, 36(4):233--240, July 2022.

\bibitem{maatouk2020}
A.~Maatouk, S.~Kriouile, M.~Assaad, and A.~Ephremides.
\newblock The age of incorrect information: A new performance metric for status
  updates.
\newblock {\em IEEE/ACM Transactions on Networking}, 28(5):2215--2228, October
  2020.

\bibitem{yang2018}
A.~Baknina, O.~Ozel, J.~Yang, S.~Ulukus, and A.~Yener.
\newblock Sending information through status updates.
\newblock In {\em IEEE ISIT}, June 2018.

\bibitem{ozel2022}
O.~Ozel, A.~Yener, and S.~Ulukus.
\newblock State amplification and masking while timely updating.
\newblock In {\em IEEE ISIT}, June 2022.

\bibitem{bastopcu2021}
M.~Bastopcu and S.~Ulukus.
\newblock Timely tracking of infection status of individuals in a population.
\newblock In {\em IEEE Infocom}, May 2021.

\bibitem{yates-survey}
R.~D. Yates, Y.~Sun, D.~R. Brown, S.~K. Kaul, E.~Modiano, and S.~Ulukus.
\newblock Age of information: An introduction and survey.
\newblock {\em IEEE Jour. Sel. Areas in Comm.}, 39(5):1183--1210, May 2020.

\bibitem{sun-survey}
Y.~Sun, I.~Kadota, R.~Talak, and E.~H. Modiano.
\newblock Age of information: A new metric for information freshness.
\newblock {\em Age of Information}, 12(2):1--224, December 2019.

\bibitem{kosta-survey}
A.~Kosta, N.~Pappas, and V.~Angelakis.
\newblock Age of information: A new concept, metric, and tool.
\newblock {\em Foundations and Trends in Networking}, 12(3):162--259, November
  2017.

\bibitem{yates2012}
S.~Kaul, R.~Yates, and M.~Gruteser.
\newblock Real-time status: How often should one update?
\newblock In {\em IEEE Infocom}, March 2012.

\bibitem{yates2019}
R.~D. Yates and S.~K. Kaul.
\newblock The age of information: Real-time status updating by multiple
  sources.
\newblock {\em IEEE Transactions on Information Theory}, 65(3):1807--1827,
  March 2019.

\bibitem{kam2018}
C.~Kam, S.~Kompella, G.~D. Nguyen, J.~E. Wieselthier, and A.~Ephremides.
\newblock On the age of information with packet deadlines.
\newblock {\em IEEE Transactions on Information Theory}, 64(9):6419--6428,
  September 2018.

\bibitem{chen2016}
K.~Chen and L.~Huang.
\newblock Age-of-information in the presence of error.
\newblock In {\em IEEE ISIT}, July 2016.

\bibitem{burke1956}
P.~J. Burke.
\newblock The output of a queuing system.
\newblock {\em Operations research}, 4(6):699--704, December 1956.

\bibitem{cassandras2008book}
C.~G. Cassandras and S.~Lafortune.
\newblock {\em Introduction to Discrete Event Systems}.
\newblock Springer, 2008.

\end{thebibliography}
\end{document}